\documentclass{article}
\usepackage{spconf,amsmath,graphicx}
\newcommand\norm[1]{\left\lVert#1\right\rVert}
\usepackage{flushend}

\title{META-OPTIMIZATION OF DEEP CNN FOR IMAGE DENOISING USING LSTM}
\name{Basit O. Alawode\textsuperscript{1}, Motaz Alfarraj\textsuperscript{2}}
\address{King Fahd University of Petroleum and Minerals,\\
	Electrical Engineering Department,\\ 
	Dhahran, Saudi Arabia.\\
	Email: \{g201707310\textsuperscript{1}, motaz\textsuperscript{2}\}@kfupm.edu.sa.}

\begin{document}
%
\maketitle
\begin{abstract}
\textbf{The recent application of deep learning (DL) to various tasks has seen the performance of classical techniques surpassed by their DL-based counterparts. As a result, DL has equally seen application in the removal of noise from images. In particular, the use of deep feed-forward convolutional neural networks (DnCNNs) has been investigated for denoising. It utilizes advances in DL techniques such as deep architecture, residual learning, and batch normalization to achieve better denoising performance when compared with the other classical state-of-the-art denoising algorithms. However, its deep architecture resulted in a huge set of trainable parameters. Meta-optimization is a training approach of enabling algorithms to learn to train themselves by themselves. Training algorithms using meta-optimizers have been shown to enable algorithms achieve better performance when compared to the classical gradient descent-based training approach. In this work, we investigate the application of the meta-optimization training approach to the DnCNN denoising algorithm to enhance its denoising capability. Our preliminary experiments on simpler algorithms reveal the prospects of utilizing the meta-optimization training approach towards the enhancement of the DnCNN denoising capability.}
\end{abstract}
\section{Introduction}
\label{sec:intro}

Most real-world images are captured in noisy environments. For instance, an image captured in a dusty environment. A noisy image $ Y \in {\rm I\!R}^{N \times M}  $ is therefore a combination of the clean image $ X \in {\rm I\!R}^{N \times M}  $ and some noise $ W \in {\rm I\!R}^{N \times M}  $. This is represented mathematically as below:  

\begin{equation}
	\label{eq:noisy_image}
	Y = X + W
\end{equation}

Although there are many models of $ W $ in literature such as poisson, salt, pepper, etc., the most commonly assumed is the Gaussian model because of its mathematical simplicity and ability to model practical data. Hence, $ W $ is typically referred to be additive, white, and Gaussian noise (AWGN) having a mean of zero and a specified variance ($ \sigma^2 $). We, therefore, desire to recover the clean image $ X $ from the noise corrupted version $ Y $. This is done with the help of image denoising algorithms. 

As images form an integral part of our everyday life, image denoising has been an active area of research for a very long time. As such, researchers have proposed several algorithms to help in the recovery of the clean image from the noise corrupted version. These algorithms can broadly be categorized into two classes, i) Classical, and ii) Neural Network (NN) based algorithms. The classical denoising algorithms take advantage of various mathematical concepts to perform denoising. Instances of the algorithms in this category are the pixel-based non-local means (NLM) \cite{buades2005}, the patch-based Block-Matching and 3D transformation (BM3D) \cite{Dabov2007a}, Collaborative Support-Agnostic Recovery (CSAR) \cite{Behzad2017}, K-SVD \cite{Aharon2006a} denoising algorithms, etc. Typically, patch-based algorithms perform denoising via transformation from the spatial domain to another domain. Transforms such as the curvelet \cite{starck2002}, wavelet \cite{daubechies1990}, contourlet \cite{do2005}, discrete cosine  (DCT) and discrete wavelet (DWT) \cite{qayyum2016} are utilized for such transformation. The majority of classical denoising algorithms have been based on image patches. This is because a patch typically carries more information about an image when compared to a single pixel.

In spite of the gradual improvements in the performance of the classical denoising algorithms over the years, the recent application of deep learning (DL) in many areas, image processing inclusive, have seen the performance of the classical algorithms surpassed by their deep learning counterparts. This can be attributed to the availability of enormous data and processing power. Neural networks (NNs) form the heart of DL. Algorithms such as Trainable Nonlinear Reaction-Diffusion (TRND) \cite{chen2017}, block-matching CNN (BMCNN) \cite{ahn2018}, generative adversarial networks (GANs) \cite{Zhiping2019}, fast feedforward NN (FFDNet) \cite{Zhang2018}, and deep feed-forward CNN (DnCNN) \cite{Zhang2017} are some of the DL-based denoising algorithms that have been proposed. Generally, these algorithms achieve better denoising performance when compared to the state-of-the-art classical algorithms. Due to the better performance of the DL-based methods, some of the classical algorithms' pipelines have been redeveloped using DL techniques. Of such are the BM3D-Net \cite{Yang2018a} and Deep-KSVD \cite{Scetbon2019a} which transform the classical BM3D and K-SVD denoising pipelines into DL-based versions respectively. These transformed versions have been shown to achieve better performance when compared with their classical counterparts. 

The DnCNN particularly takes advantage of advances in DL regularization techniques such as batch normalization \cite{Szegedy2015a} and residual learning \cite{He2016} to achieve impressive denoising performance. However, its deep architecture resulted in a huge set of trainable parameters. 

Meta-optimization is a training approach of enabling algorithms to learn to train themselves by themselves. Training algorithms using meta-optimizers have been shown to enable algorithms achieve better performance compared to the classical gradient descent-based training approach. In this work, we investigate the application of the meta-optimization training approach to the DnCNN denoising algorithm to enhance its denoising capability.

The remainder of this paper is organized as follows: In Section \ref{sec:related_work}, we present the detail review of the DnCNN algorithm, its strengths and associated challenges. We also provide an overview of meta-optimization in the same Section. In Section \ref{sec:methodology}, we present the methodology of applying meta-optimization method to the DnCNN algorithm. The results and conclusion are discussed in Sections \ref{sec:results} and \ref{sec:conclusion} respectively.

\section{Related Work}
\label{sec:related_work}

\subsection{The DnCNN Denoising Algorithm}

\begin{figure*}
	\centering
	\includegraphics[width=\textwidth]{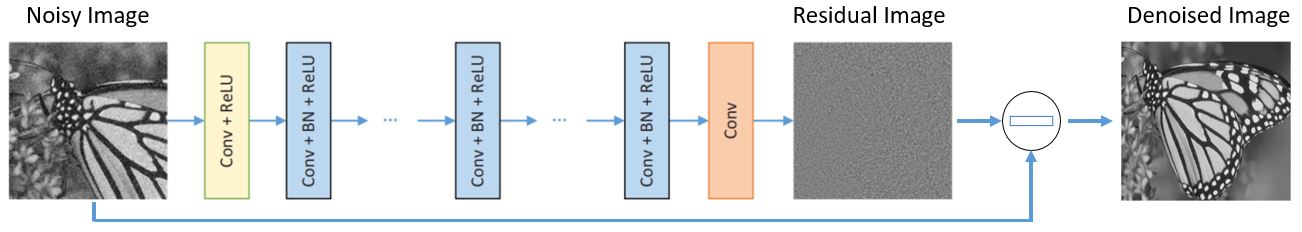}
	\caption{The DnCNN Network Architecture}
	\label{fig:dncnn}
\end{figure*}

The basic architecture of the DnCNN algorithm \cite{Zhang2017} is as shown in the Figure \ref{fig:dncnn}. The network is composed of 17 convolution layers. All layers except the output layer is activated with a Rectified Linear Unit (ReLU) \cite{nair}. Its deep structure enabled it to gradually separate the image from its noisy observation. This gave it its denoising power over other architectures. Batch normalization \cite{Szegedy2015a} is applied to all hidden convolution layers to address the problem of internal co-variate shift which ensued as a result of stacking several layers of convolution. Instead of learning the clean image, the residual image is learned. The residual image is essentially the noise in the image. Once learned, the noise is then removed from the noisy image to get the denoised image. This concept is known as residual learning \cite{He2016}. The adoption of the residual learning strategy to learn the noise and the incorporation of batch normalization helped in speeding up training as well as boosting the model’s overall performance. 

The DnCNN, owing to its deep architecture and generalization ability, was able to surpass most of the denoising algorithms in performance. However, its deep architecture resulted in a CNN model with a huge set of trainable parameters. 

\subsection{Meta-Optimization Overview}

Most NNs are trained with the help of classical gradient descent-based (GD-based) optimization algorithms such as SGD, Adam, RMSProp, etc. \cite{dogo2018}. The hyperparameters of these optimizers are hand-selected based on the domain expertise of the network designer. This imply that the performance of the NN will depend on the choice of the hyperparameters of the optimizer used in training it. Is it possible to train optimizers similar to how networks are trained such that the trained optimizers would be able to optimize similar networks on which it was trained? The answer to this question lies in the concept of meta-learning otherwise known as “learning to learn”. 

With meta-optimizers, there is little interference by the designer on how the network trains itself. Since the optimization of the network is determined by another trained network (a meta-optimizer), the achievement of better performance when compared to the handcrafted GD-based optimizers is not farfetched.

The concept of meta-learning has been around for a long time \cite{pratt1998} but it became an important part of artificial intelligence (AI) in 2016 when Lake et al. argued for its importance \cite{lake2016}. There are several branches of meta-learning \cite{Hospedales2020} of which meta-optimization is an integral part. The general concept of designing a meta-optimizer is to develop a trainable optimizer capable of taking advantage of the training history (momentum) of the network they are optimizing. This implies that the meta-optimizer must possess some form of memory and evolve over the training time of the network it optimizes. As a result, reinforcement learning (RL) and variants of recurrent NN (RNN) have been used by researchers to develop these meta-optimizers because of their history retaining nature. The RL approach represents a meta-optimizer as a policy. The meta-optimizer is then learned using guided policy search \cite{Li2016,Daniel2016}. 

Due to the success of RNN in sequence tasks, RNN and its variants such as the Long Short-Term Memory (LSTMs) have seen more applications in the design of meta-optimizers \cite{Li2016,Ravi2017,Metz2018}. Particularly, in 2016, Andrychowicz et al. \cite{Andrychowicz2016} proposed a meta-optimizer using LSTMs. The memory structure of the LSTM allows the meta-optimizer to keep track of past training steps applied to the base network being trained in order to predict future optimization updates. The LSTM-based optimizer is itself trained using GD-based optimizers. With no interference by the designer, the meta-optimizers typically outperform the classical optimizers on the tasks for which they are trained.

\section{Methodology}
\label{sec:methodology}

In this work, we apply meta-optimization to extend the denoising capability of the DnCNN algorithm. We approach this in 2 stages. The first is to redevelop the DnCNN algorithm and train it using the standard classical GD-based training methods. We then proceed to investigate the use of meta-optimizers on the DnCNN to improve upon its performance.

\subsection{Reproducing the DnCNN Algorithm}
\label{sec:reproduce_dncnn}

The DnCNN algorithm as shown in Figure \ref{fig:dncnn} was developed. We stacked 17 CNN layers with batch-normalization and ReLU activation layers added in-between each activation layer. From the Figure, it can be observed that the input CNN layer is not batch-normalized. This is because batch-normalization is meant to correct internal co-variate shifts which can only occur in the hidden layers. The input is our noisy image while the output of the CNN layers is the noise in the image called the residual image. This is then subtracted from the input image to obtain the denoised image.

\subsection{Meta-optimizing the DnCNN Algorithm}

\begin{figure*}
	\centering
	\includegraphics[width=\textwidth]{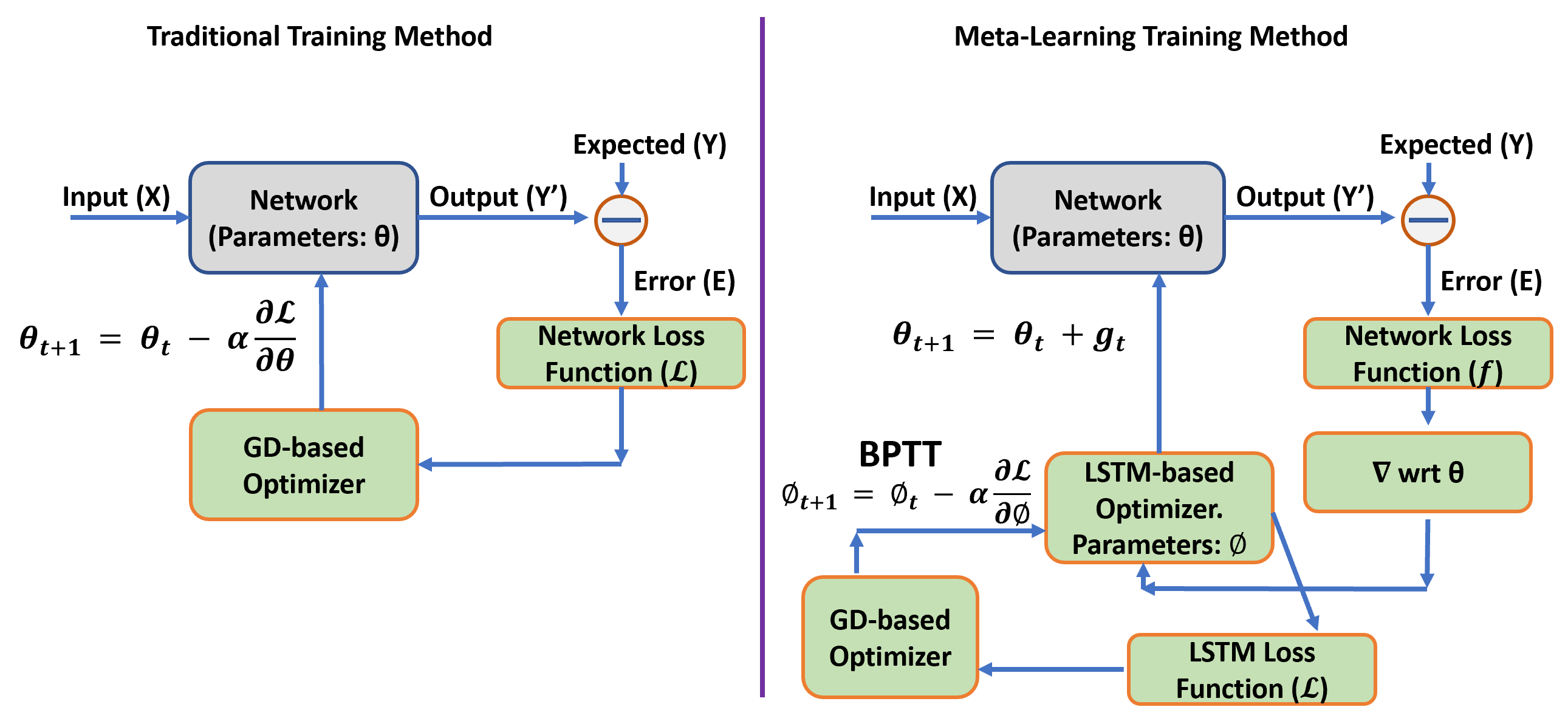}
	\caption{Traditional vs Meta-Learning Training Methods}
	\label{fig:meta}
\end{figure*}

The meta-optimization technique we have employed in this work is similar to that proposed in \cite{Andrychowicz2016}. To better explain the meta-optimization steps, we show in the Figure \ref{fig:meta} an oversimplified comparison between the traditional and the meta-learning training methods. In the traditional training methods (gradient descent, GD-based), we pass the input to the network and obtain the output. The error which is the difference between the predicted output and the expected output is then computed. Based on this, a loss function is obtained which is then used by the GD-based optimizer to update the parameters of our network. This is typically done via backpropagation.

The meta-learning training method takes this further. We utilized the gradients of the parameters obtained from the computed loss function to train the meta-optimizer. It is important that the optimizer be a trainable network having memory capability such as RNN (for instance, LSTMs)  and Transformers \cite{Vaswani2017}. The meta-optimizer we have used here is an LSTM-based one. 

To implement the LSTM-based meta-optimizer for the DnCNN algorithm which consists mainly of CNNs, the meta-optimizer would be trained on how to optimize CNN-based networks. As such, we trained the meta-optimizer on a subnetwork of the DnCNN called the base-network by minimizing its error using the updates from the meta-optimizer. Similar to GD-based optimizers, the input to the meta-optimizer is the gradients of the parameters of the base-network and the output is the update to be applied on the parameters that would minimize the error. The difference is that the meta-optimizer is trained to output parameter updates which would yield a better performance for the base-network based on the base-network's training history (that is, training memory). More details on this meta-optimization training approach can be found in \cite{Andrychowicz2016}. 

\section{Results and Discussions}
\label{sec:results}

The performance of denoising algorithms are usually measured and compared subjectively and objectively. In this work, we have focused more on the objective measures. These are image quality assessment (IQA) algorithms such as the peak signal to noise ratio (PSNR) and the structural similarity index (SSIM) \cite{Al-Najjar2012, Wang2002a}. We also discuss the visual performance of the algorithms as a form of subjective measure.

\subsection{Training and Testing Data}

Similar to the authors of the DnCNN algorithm, we have utilized a set of 400 images of size 180 × 180 for training from the Berkeley Segmentation Dataset (BSD) \cite{chen2017, bsd2001}. From the images, $ 128 \times 1600 $ clean overlapping patches of size $ 40 \times 40 $ were cropped out to generate enough training data. WGN of known variance ($\sigma $) in the range [$5$, $90$] were then added to the patches to generate the noisy images. For testing, 80 widely used natural images, also from the BSD dataset have been set utilized. It is to be noted that the testing images were not included in the training dataset.

\subsection{Network Training}

The DnCNN is trained using the noisy patches as the input and the corresponding clean patches at the output. Training is done for 50 epochs with a decaying learning rates using the Adam optimizer \cite{kingma2015} with default parameters. The mean squared error (MSE) loss function has been used.

The meta-optimizer implements a 2-layer LSTM network with 20 hidden units per layer. To train the DnCNN using the meta-optimizer, the LSTM-based optimizer is first trained on how to optimize CNN-based networks. This is done by training a smaller 5 layer DnCNN subnetwork as the base-network. For each training epoch, the MSE error of the base-network is used to update the parameters of the LSTM-based meta-optimizer. Training the optimizer is achieved using the truncated backpropagation through time (TBPTT) optimized with the Adam optimizer using 100 epochs. Finally, the trained meta-optimizer is applied to train the full DnCNN network.

\subsection{Compared Methods}

We compare the DnCNN trained using the meta-optimizer with several state-of-the-art denoising algorithms. These algorithms are the NLM \cite{buades2005}, KSVD \cite{Aharon2006a}, and the BM3D \cite{Dabov2007a} algorithms. We also compare it with the DnCNN \cite{Zhang2017} algorithm trained using the Adam optimizer.

\subsection{Hardware Requirements and Code Development}

All the training and testing codes for the DnCNN and the meta-optimizer were written using the Python-based DL Keras \cite{keras} library with the TensorFlow library as the backend. Training was performed using 2 Nvidia GTX 1080 Ti GPUs running on a Linux-based cluster with a pre-allocated 32GB of RAM. Testing was performed on an Intel Core i7 8th Gen CPU with a 16GB of RAM running a Windows 10 operating system. 

It should be noted that the while we got a head start with the DnCNN algorithm here \cite{cszndncnn}, it was completely rewritten to achieve the desired result. The code for the meta-optimizer has also been completely written from scratch. For comparison, the release NLM, BM3D, and K-SVD algorithms' MATLAB codes have been utilized. Although, there were modifications to obtain our desired results, we did not tamper with the base code of these algorithms.

\subsection{Objective Comparison}

The average PSNR and SSIM results of the different algorithms on the BSD80 dataset are as shown in the Table \ref{t:d_var}. It can be observed that the DnCNN trained on the GD-based Adam optimizer performed better than all the other state-of-the-art algorithms for all considered noise variances. The testing images are of varying sizes, resolutions, and details. As such, the result presented in the Table is a fair representation of the denoising capability of the different algorithms.

Although we expect that the DnCNN trained using the developed meta-learning should perform better than the GD-trained algorithm, but, as of the time of writing this report, we have not been able to successfully apply this to the DnCNN to improve upon its result. This is evident in the meta-trained DnCNN result shown in the Table \ref{t:d_var}. Solving this challenge was a major part of this work. However, a positive note on the meta-trained DnCNN result is that the algorithm did not result in degraded performance as can be observed in the Table.

\begin{table*}[h]
	\centering
	\caption{Average Denoising Performance (80 Test Images)}
	\begin{tabular}{ |c|c|c|c|c|c|c|c| }
		\hline
		\hline
		Noise Variance & SNR & Noisy & NLM & KSVD & BM3D & DnCNN & Our DnCNN \\
			$ \sigma $		   & (dB) &  &  &  &  & (GD-trained) & (meta-trained) \\
		\hline
		\hline 
		5&26.80 &34.15 &35.95 &37.38 &37.70 &\textbf{38.07} &34.21 \\
		& &0.89 &0.94 &0.96 &0.97 &\textbf{0.98} &0.95 \\
		\hline
		10&20.78 &28.13 &31.88 &33.15 &33.49 &\textbf{34.03} &28.24 \\
		& &0.71 &0.88 &0.92 &0.92 &\textbf{0.96} &0.84 \\
		\hline
		15&17.26 &24.61 &29.67 &30.93 &31.22 &\textbf{31.90} &24.78 \\
		& &0.57 &0.83 &0.87 &0.88 &\textbf{0.94} &0.72 \\
		\hline
		20&14.76 &22.11 &28.20 &29.45 &29.75 &\textbf{30.45} &22.33 \\
		& &0.47 &0.78 &0.83 &0.85 &\textbf{0.95} &0.61 \\
		\hline
		25&12.82 &20.17 &27.11 &28.35 &28.64 &\textbf{29.26} &20.45 \\
		& &0.39 &0.74 &0.80 &0.82 &\textbf{0.91} &0.52 \\
		\hline
		30&11.24 &18.59 &26.25 &27.47 &27.79 &\textbf{28.45} &18.95 \\
		& &0.34 &0.71 &0.77 &0.79 &\textbf{0.89} &0.45 \\
		\hline
		40&8.74 &16.09 &24.92 &26.13 &26.31 &\textbf{27.31} &16.62 \\
		& &0.25 &0.64 &0.71 &0.74 &\textbf{0.86} &0.34 \\
		\hline
		50&6.80 &14.15 &23.93 &25.09 &25.35 &\textbf{26.28} &14.89 \\
		& &0.20 &0.59 &0.67 &0.70 &\textbf{0.83} &0.27 \\
		\hline
		60&5.22 &12.57 &23.16 &24.26 &24.60 &\textbf{25.65} &13.54 \\
		& &0.16 &0.54 &0.63 &0.67 &\textbf{0.81} &0.22 \\
		\hline
		70&3.88 &11.23 &22.52 &23.58 &23.97 &\textbf{24.98} &12.23 \\
		& &0.13 &0.50 &0.60 &0.64 &\textbf{0.78} &0.19 \\
		\hline
		80&2.72 &10.07 &21.99 &22.99 &23.44 &\textbf{24.59} &11.60 \\
		& &0.11 &0.47 &0.58 &0.62 &\textbf{0.78} &0.16 \\
		\hline
		90&1.70 &9.04 &21.53 &22.51 &22.99 &\textbf{24.03} &10.89 \\
		& &0.09 &0.43 &0.56 &0.60 &\textbf{0.76} &0.14 \\
		\hline	
		\hline
	\end{tabular}
	\label{t:d_var}
\end{table*}

\subsection{Subjective Comparison}
\label{sec:visual_comp}

Objective measures are generally not enough to completely have a conclusion on an image processing algorithm \cite{Al-Najjar2012}. For this reason, we subjectively (visually) compare the algorithms. We have presented the denoised images for noise variance $\sigma = 15 $ in Table \ref{t:visual_comp}. We can observe that details are lost in the NLM and K-SVD denoised images. The BM3D and DnCNN (GD-trained) have similar visual performance. However, the BM3D slightly lost some of its details as compared to the DnCNN especially in the sand region of the denoised image. This also confirms the superiority of the DnCNN algorithm. Also, our DnCNN (meta-trained), while achieving a slight denoising result still contains a huge amount of noise. This is currently being investigated.

\begin{table*}[h]
	\caption{Visual Comparison at $\sigma = 15 $} 
	\label{t:visual_comp} 
	\begin{tabular}{c c c c}
		\hline
		\hline
		(a) - Original & (b) - Noisy & (c) - NLM  & (d) - K-SVD \\
		\includegraphics[width= 0.22\textwidth]{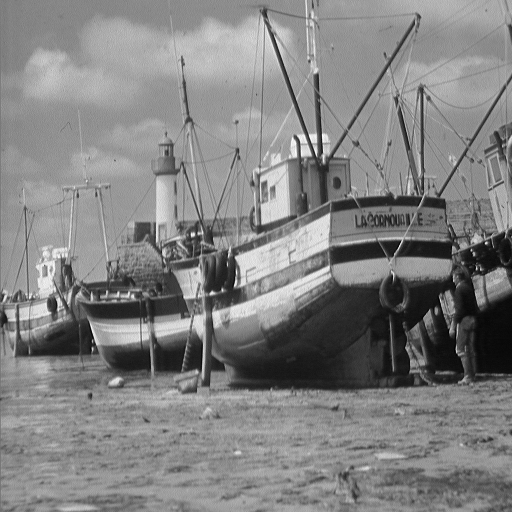} &	
		\includegraphics[width= 0.22\textwidth]{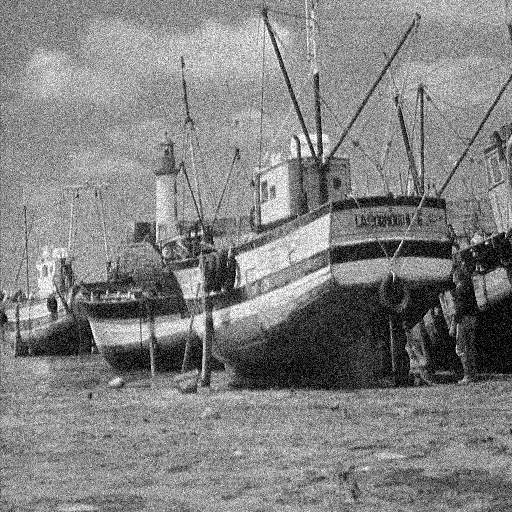} 
		&
		\includegraphics[width= 0.22\textwidth]{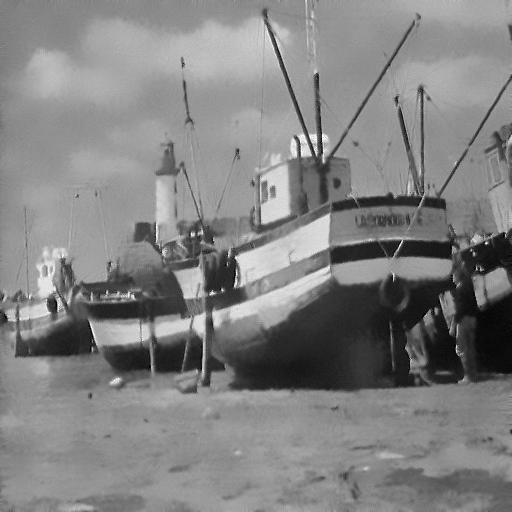} & 
		\includegraphics[width= 0.22\textwidth]{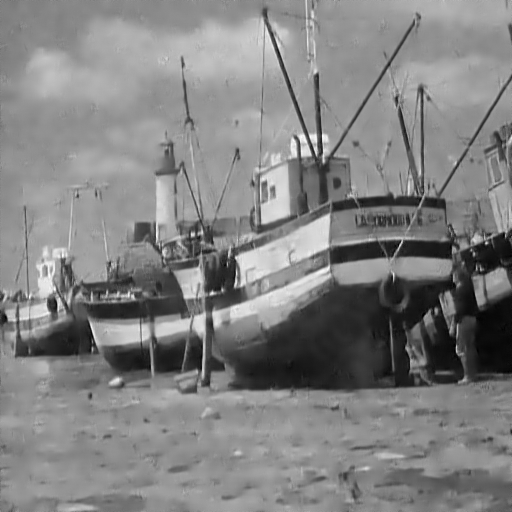}
		\\
		\hline
		(e) - BM3D & (f) - DnCNN (GD-trained) & (g) - DnCNN (meta-trained) \\	
		\includegraphics[width= 0.22\textwidth]{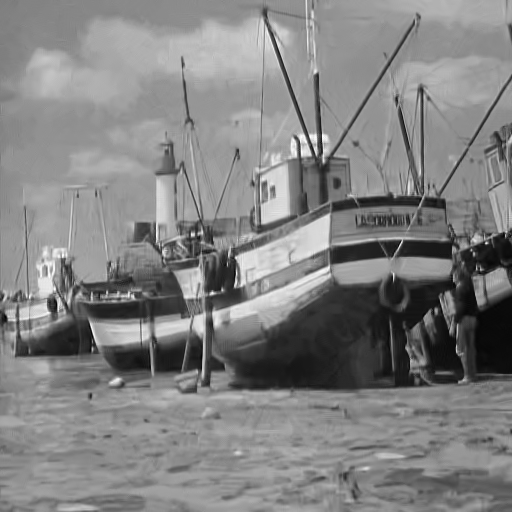}
		&
		\includegraphics[width= 0.22\textwidth]{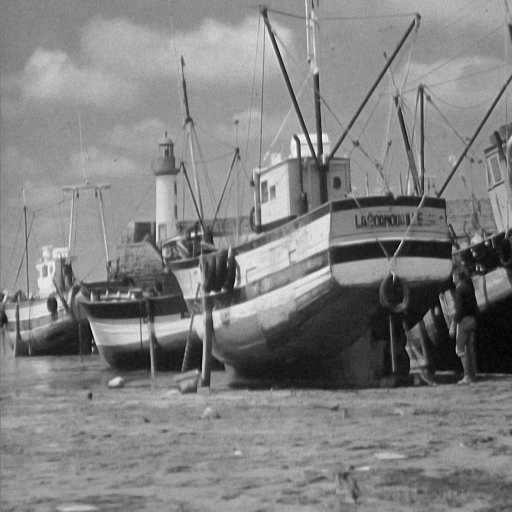} &
		\includegraphics[width= 0.22\textwidth]{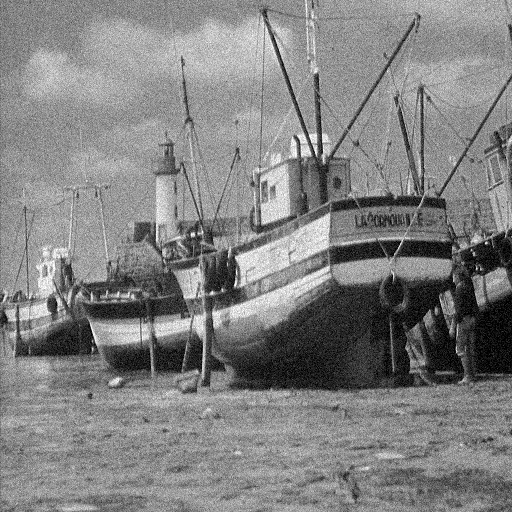}
		\\
		\hline
		\hline
	\end{tabular}
\end{table*}

\subsubsection{Denoising Time}
\label{sec:time}

Another important consideration for any image restoration method is the duration. Here, we compare the average denoising time of the different algorithms. This is as shown in the Table \ref{tab:denTimeCompare}. These times, as we have observed through our experiments, cannot be accurately predicted as they vary from one experiment to the next. They depend largely on many factors from the PC such as the available RAM, internal PC temperature, number of opened applications, etc. However, we have presented the average time observed throughout our various experiments. From the Table, we can observe that both DnCNNs (GD trained and meta-trained) have similar and least denoising time compared to the other algorithms. 

On the average, training the DnCNN with GD-based optimizer takes about 230 minutes (which can sometimes be as high as 300 minutes). Training the DnCNN with the meta-optimizer takes about similar time as the GD-based optimizer. However, the meta-optimizer has to be trained for about 185 minutes on the average. 

\begin{table}[!htbp]
	\centering
	\caption{Average Denoising Time}
	\begin{tabular}{l|c}
		\hline
		\textbf{Algorithm} & \textbf{Time (seconds)} 
		\\
		\hline
		& \\
		NLM & 15  \\
		& \\
		K-SVD & 60  \\
		& \\
		BM3D & 30  \\
		& \\
		DnCNN (GD-trained) & 15 \\
		& \\
		DnCNN (meta-trained) & 13 \\
		\hline
		\hline
	\end{tabular}
	\label{tab:denTimeCompare}
\end{table}

\subsection{Meta-optimizer on Other Networks}

In spite of our failure to successfully apply the meta-optimizer on the DnCNN, we have performed experiments to show that the performance of networks can be improved by training them using meta-optimizers. In these experiments, we applied the optimizer on simpler networks. These experiments are similar to those performed by the authors in \cite{Andrychowicz2016} and are as presented below.

\subsubsection{Quadratic Functions}

Here, we trained the meta-optimizer on a simple class of synthetic 10-dimensional
quadratic functions. Particularly, we minimized functions of the form shown \eqref{eq:quad}.

\begin{equation}
	\label{eq:quad}
	f(\theta) = \norm{P\theta - y}^2_2
\end{equation}

We performed the minimization for different $ 10\times10 P $ matrices and $ 10 $-dimensional vectors $ y $ whose elements are Gaussian independent and identically distributed (IID) random variables. The optimizer is trained by optimizing random functions from this family. Since this is LSTM-based, we unrolled for 20 steps and each function was optimized for 100 epochs. 

\begin{figure}[h]
	\centering
	\includegraphics[width=0.45\textwidth]{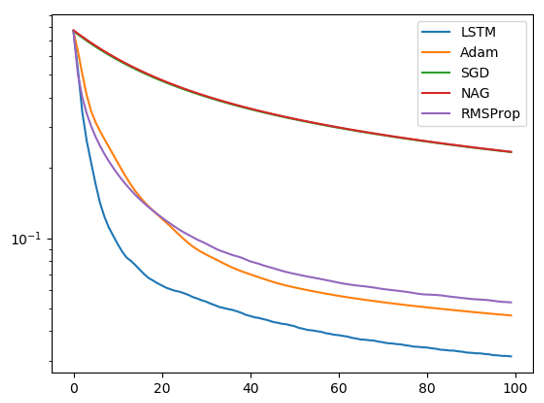}
	\caption{LSTM-Based Optimizer on Quadratic Data (NN)}
	\label{fig:quadratic}
\end{figure}

We tested on newly sampled functions from the same distribution and then compared the trained optimizer performance with several other GD-based optimizers. This is as presented in the Figure \ref{fig:quadratic}. It is clear that the meta-optimizer outperforms the other baseline optimizers. 

\subsubsection{Dense-NN on MNIST}

\begin{figure}[h]
	\centering
	\includegraphics[width=0.45\textwidth]{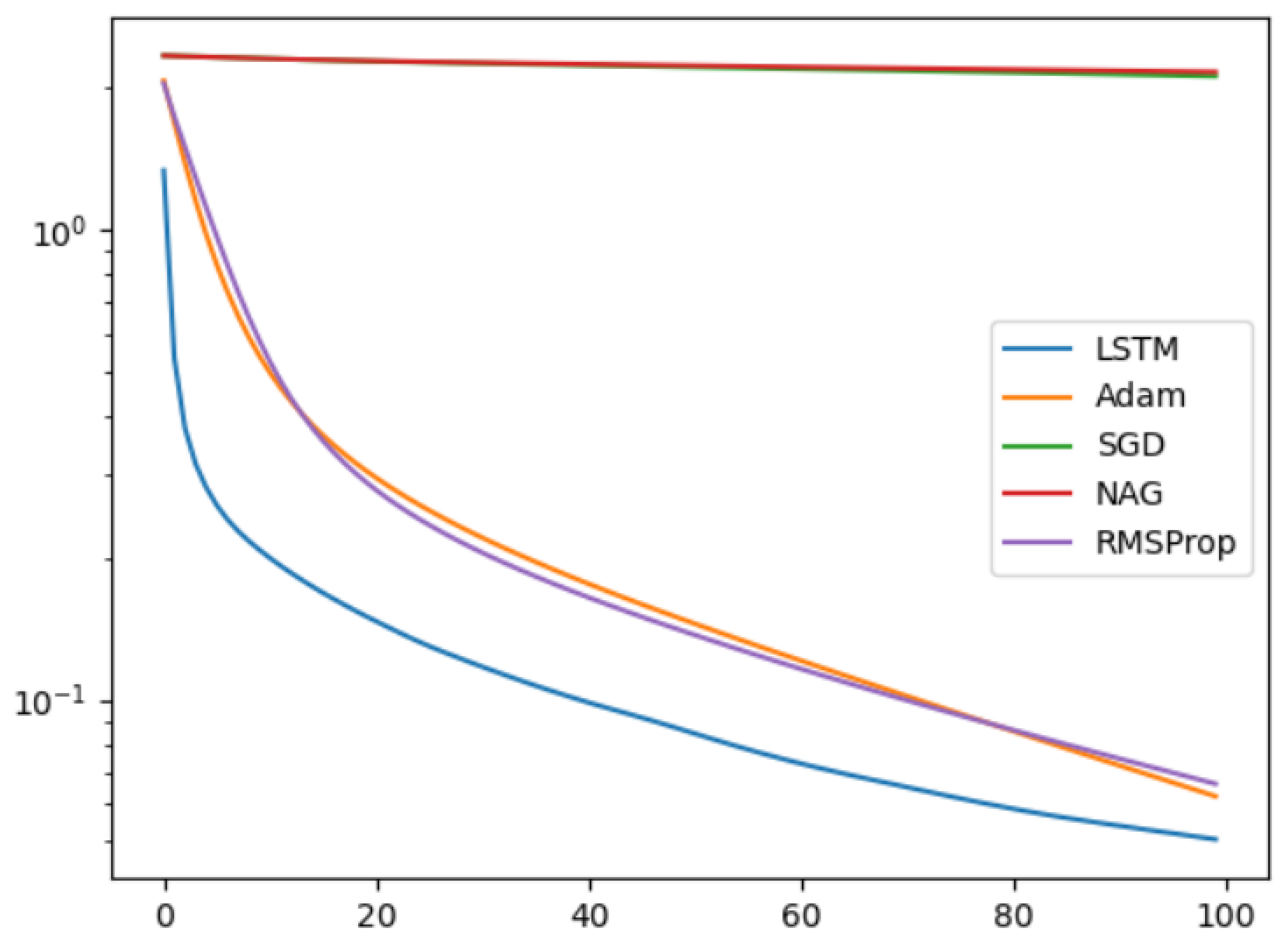}
	\caption{LSTM-Based Optimizer on MNIST Dataset (NN)}
	\label{fig:mnist}
\end{figure}

We moved a step further by investigating the meta-optimizer for multilayer dense NNs. For this, we utilized the MNIST handwritten digit dataset trained for classification. We trained the meta-optimizer using a 2-hidden layer dense NN with 20 neurons per layer. As this is a handwritten digit recognition task, the output layer contains 10 neurons with a sigmoid activation. Testing is performed on 2 networks - 1.) a similar network configuration as the training and 2.) another network with 40 neurons instead of 20. Both experiments yielded similar performance as shown in Figure \ref{fig:mnist}. Clearly, the LSTM-based meta-optimizer performed better than the other classical optimizers.

\subsection{Future Work}

As of the time of compiling this report, we have been able to apply the meta-optimizer on the DnCNN. However, we have not been able to yield the desired result. This can be attributed to the manner in which CNN-based network filter weights are trained. We believe that with more effort, the desired objective would be achieved.

\section{Conclusion}
\label{sec:conclusion}

In this work, we have applied an LSTM-based meta-optimizer to the deep feedforward convolutional neural network (DnCNN) denoising algorithm. The choice of the DnCNN stems from the fact that deep learning image denoising techniques outperformed many state-of-the-art classical image denoising algorithms. In particular, we showed through experiments on several images how the DnCNN performed better than many benchmark classical denoising algorithms such as the non-local means (NLM), block-matching 3D filtering (BM3D), and the K-singular value decomposition (K-SVD) algorithms. We seek to extend the performance of the DnCNN further by training it with a learned LSTM-based meta-optimizer, instead of the traditional gradient descent-based optimization approaches. Our current result of the meta-trained DnCNN did not perform as expected. However, a positive note on the meta-trained DnCNN result is that the algorithm did not result in degraded performance. Further experiments on using a learned meta-optimizer on simple function approximation networks and dense neural networks revealed that the performance of algorithms can be further extended by training them with meta-optimizers. This approach of training the DnCNN to extend its denoising performance is currently being investigated.

\section{Acknowledgment}

We gratefully acknowledge the support from the KAUST supercomputing lab for providing us with remote access to the GPUs that were used in this work.


\bibliographystyle{IEEEtran}
\bibliography{/Users/BASTECH-LPC/Documents/Mendeley/KAUST,/Users/BASTECH-LPC/Documents/Mendeley/KAUST-Meta-Learning}

\end{document}